# Multi-timescale Event Detection in Nonintrusive Load Monitoring based on MDL Principle

Bo Liu, Jianfeng Zhang, Wenpeng Luan*, *Senior Member*, Zishuai Liu, *IEEE*, Yixin Yu, *Life Senior Member*, *IEEE*

***Abstract*—Load event detection is the fundamental step for the event-based non-intrusive load monitoring (NILM). However, existing event detection methods with fixed parameters may fail in coping with the inherent multi-timescale characteristics of events and their event detection accuracy is easily affected by the load fluctuation. In this regard, this paper extends our previously designed two-stage event detection framework, and proposes a novel multi-timescale event detection method based on the principle of minimum description length (MDL). Following the completion of step-like event detection in the first stage, a long-transient event detection scheme with variable-length sliding window is designed for the second stage, which is intended to provide the observation and characterization of the same event at different time scales. In that, the context information in the aggregated load data is mined by motif discovery, and then based on the MDL principle, the proper observation scales are selected for different events and the corresponding detection results are determined. In the post-processing step, a load fluctuation location method based on voice activity detection (VAD) is proposed to identify and remove the unreasonable events caused by fluctuations. Based on newly proposed evaluation metrics, the comparison tests on public and private datasets demonstrate that our method achieves higher detection accuracy and integrity for events of various appliances across different scenarios.**

## I. Introduction

NONINTRUSIVE load monitoring (NILM) [1] is a cost-effective way to get itemed energy consumption of individual appliances by analyzing the aggregated load data. Its implementation will enhance the observability of power consumption on the demand side and facilitate the trace and analysis of carbon footprint, so as to support energy conservation and emission reduction. According to statistics [2], customer power savings driven by informative feedback on power usage details can reach 5% - 20%.

With the ultimate goal to realize load identification and disaggregation, NILM comprises of two general processes—learning and inference [3]. The learning process aims at deriving the appliance model and the parameters of the NILM algorithm, while the inference process identifies and infers the operation states and the power consumption of appliances according to the aggregated load data. Furthermore, the existing NILM methods can be classified as event-based and non-event-based ones based on whether an event detection step is performed before the learning and inference process [4][5]. Load event (will be referred as “event” later) corresponds to the transition process of any two defined operating states of appliances, which is represented on the aggregated load power profile as a transient segment between two adjacent distinctly different steady-state segments, as shown schematically in Fig. 1. Different from the non-event-based methods which require inference on all samples of the data, the event-based NILM methods take the load events as the objects of learning and inference, which has the significant advantages of low algorithm complexity and high operation efficiency [6]. Therefore, the event-based solutions are the mainstream choice for NILM technology implementations.

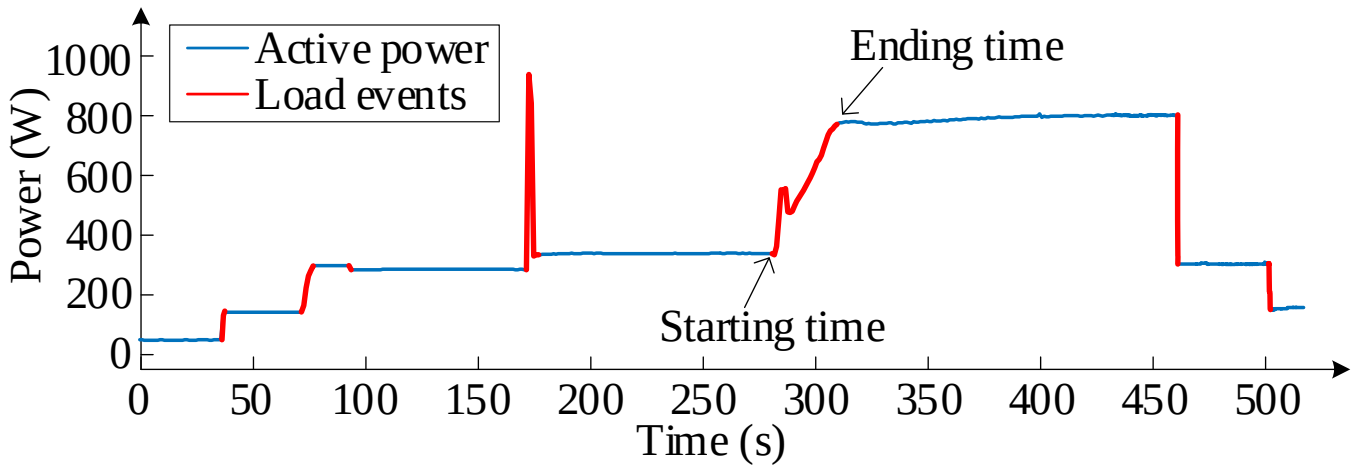

Fig. 1. Load event diagram.

The general flow of event-based NILM method is shown in Fig. 2. As can be seen, load event detection is the key step of the process, through which the transient and steady-state segments of the aggregated load data can be divided, thus the transient and steady-state load signatures of the appliances can be extracted. The load signatures extracted around the event can be either a feature variable obtained by differentiating the steady-state signals before and after event, or an additional “macroscopic” signature [7] such as the shape of the signal time series corresponding to the event itself. In the learning and inference process, load signatures are used as the feature attributes to characterize event samples. Since events are the bridge of appliance operation state transition, on one hand, the set of appliance operation states can be identified by clustering the load event samples [8], and the switching relationship between different operation states of appliances can be determined by means of event sequence pattern mining [9] or event pairing/matching [10], thus eventually to derive the complete appliance state model. On the other hand, the classification of event samples enables the identification of the operation states of appliances and the energy disaggregation. Therefore, the correct and complete detection of events (determining the starting and ending times of events as shown in Fig. 1) will provide accurate and descriptive event samples, facilitate the subsequent feature-based or distance-based event clustering and classification, and thus influence the final load identification and disaggregation performance.

This work was supported by the National Nature Science Foundation for Young Scholars of China (No. 52107120) and the Smart Grid Joint Funds of the National Natural Science Foundation of China (No. U2066207). (*Corresponding author: Wenpeng Luan*).

The authors are with the School of Electrical and Information Engineering, Tianjin University, Tianjin 300072, China (e-mail: liubo@tju.edu.cn; zhangjianfeng@tju.edu.cn; wenpeng.luan@tju.edu.cn; zishuailiu@tju.edu.cn; yixinyu@tju.edu.cn;).

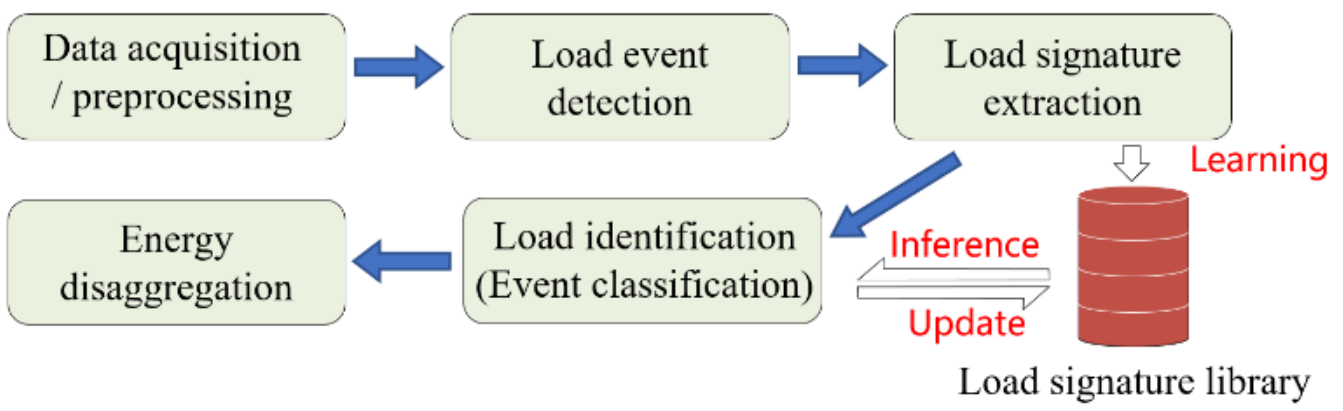

Fig. 2. The general flow of event-based NILM method.

Focusing on the load event detection, which is the essential and fundamental step for the event-based NILM, this paper proposes a novel multi-timescale load event detection method based on the minimum description length (MDL) principle, so as to improve the NILM solution's adaptive capability for events with different time scales and various shape complexity in differentiated operation scenarios. Compared with existing methods including our previous work [4], which will be reviewed in Section II, our major contributions are as follows:

1) An event detection scheme with variable-length sliding windows is proposed, which combines the trend analysis of event power time series to "observe" and characterize the different appliance transients from different time scales, so as to detect the long-transient events of appliances completely, especially for the complex appliances.

2) To address the situation that event detection results may compete at different observation scales, the MDL-based motif discovery method is used to mine the context information in the aggregated load data, and the proper observation scales are selected for different events, thus the results that correspond to the complete appliance transient process are determined.

3) A load fluctuation segment location method based on voice activity detection (VAD) technique is proposed in the event post-processing step to remove the unreasonable events detected due to the impacts of load fluctuations.

The rest of this paper is organized as follows. Section II summarizes the related work and existing problems. Section III details the proposed method. Based on newly proposed evaluation metrics, the comparison test results between the proposed method and two state-of-the-art methods on private and public data sets are presented and analyzed in Section IV. Section V summarizes the paper and gives outlook.

## II. Related Work and Existing Problems

According to the review in our previous work [4], the existing event detection methods can be divided into four categories: expert heuristic, statistical-probability model based, steady-state signal clustering, and pattern matching. Among them, the effectiveness of pattern matching methods heavily depends on the accuracy of the event templates used for matching, which often need to be obtained in a supervised manner or acquired by other event detection methods [12][13]. Therefore, a comparative analysis against this class of methods will not be conducted in this paper. In fact, no matter which of the above technology paths it follows, any single event detection approach often struggles to guarantee robustness [4] [5] [14]. This is due to the complex and diverse power consumption patterns of appliances and the presence of background noise in realistic scenarios, which pose great challenges for event detection.

As shown in Fig. 1 and Fig. 3, the challenges are manifested in the diversity of events in time scale and shape complexity. However, there is one fundamental shape characteristic of events that do not change with time scales, that is, the power waveform curve of the event transient process has an obvious and consistent upward (corresponding to on event) or downward trend (corresponding to off event).

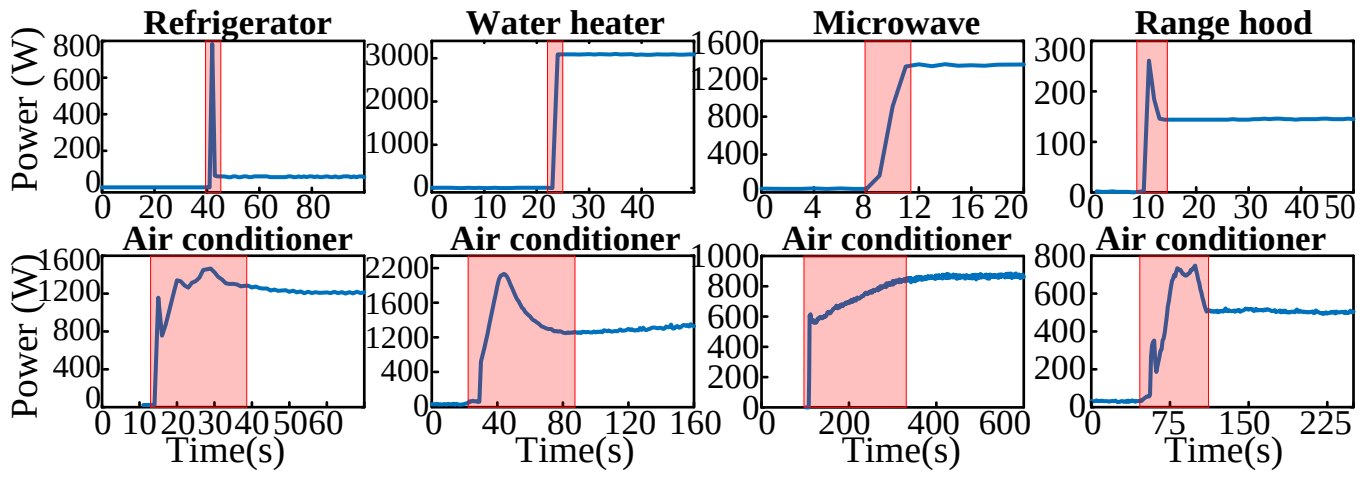

Fig. 3. Examples of active power waveforms of switching-on events

Therefore, the expert heuristic methods directly design the corresponding judgment rules according to the geometric features of events. For example, the edge detection algorithm proposed by Hart *et al*. earlier [1] classifies the samples with power variations greater than a certain threshold from adjacent samples into the event transient segments. The zero-crossing detection of the second-order differential power signal in [15] are also for the similar purpose. To avoid the influence of accidental disturbance data on the detection of the overall signal change trend of events, many scholars have adopted the sliding window based method to emphasize the signal trend on a larger time scale by calculating the overall characteristics of the signal in the window (average value [5], variance [16], etc.). The statistical-probability model-based methods use the data in the sliding window to calculate the relevant statistics, and determine whether the probability distribution of the data has changed by means of likelihood ratio [17], goodness of fit test [18], cumulative sum [19], etc., to achieve event detection. The third category of event detection methods perform clustering analysis (mean shift [20], DBSCAN [21]) on the data within the detection window, and use the transitions between adjacent steady-state signal clusters to determine the location of events.

Obviously, these window-based methods all face the problem how to set the window length. For a sliding window of arbitrary fixed length, if load events with larger time scale are encountered, it is likely that the transient segments cannot be detected completely and it is easy to incorrectly detect one event as multiple events. From another perspective, arbitrarily increasing the window length will aggravate the delay effect of the window, and tend to merge the near-simultaneous events with intervals shorter than the window length. In addition, for the method based on steady-state signal clustering, too large a window length is also likely to trigger the class imbalance problem of samples.

In order to detect the transient process of events as completely as possible, our previous research [4] proposed a two-stage adaptive event detection method. According to the different geometric features of step-like and long-transient (non-step-like) events, the improved edge detection method and the window-based detection method combining moving average and moving t-test are together used to detect events in the two stages, and the complete power trend of the event segment is also taken into account. However, a fixed window length setting can lead to inconsistent detection of the transient processes generated by the same appliance when used at different moments. For this problem, [5] adopts the following treatment: when the window-based event detection is finished, two consecutive peak or valley points with time span less than a certain threshold and near-zero amplitude are searched in the 1st derivative of the aggregated data, and then multiple events between these two points are combined into one event; however, the fixed threshold parameter limits its practical performance. Green *et al* [15] uses a multiscale median filter banks to separate the power data on different time scales, and detects the events of different scales in the medianed stream and residual stream respectively. However, it is not clear how to determine the scales of the filter bank and how to make a choice when there are contradictions between the events under different filter scales.

For these existing event detection methods, the key problem encountered can be attributed to the fact that the time-related parameters cannot be adaptive to events with different time scales and various shape complexity. In addition, most methods can only provide a single change point corresponding to the event on the time axis, and cannot accurately give the starting and ending times of the event transient segment.

Besides, the event detection is also challenged by the interference from background fluctuation or noise. Selecting a suitable detection threshold can only avoid mis-detecting fluctuations as events to a limited extent [4]. Many scholars also adopt pre-processing methods of filtering, such as median filtering [16] [18], mean filtering [21], Gaussian filtering [14]. However, these filtering methods are all global filtering for the transient and steady-state segments without any difference. On one hand, filtering may cause distortion and even destroy the shape features of the event transient power waveform, such as directly fusing the power waveforms of near-simultaneous events incorrectly. On the other hand, some appliances (e.g., electronic devices) have more complex power fluctuation characteristics during operation, and the conventional single filter is unable to adapt to the signal characteristics of different fluctuation segments, which makes it difficult to completely filter out the fluctuations. In addition, the event detection scheme with the post-processing of Savitzky-Golay filtering proposed in [5] does not filter out pseudo-events well enough and suffer from the problem of false alarms.

In order to overcome the problem that the existing methods cannot automatically determine the appropriate lengths of observation windows or related parameters for different load events and are susceptible to background load fluctuations or noise, this paper studies on the novel event detection method to realize time-related parameter adaptation and withstand the interference of fluctuation.

## III. METHODOLOGY

Based on the previously proposed adaptive two-stage event detection framework [4], this paper further proposes the MDL-based multi-timescale load event detection method, as shown in Fig. 4. Among them, the steps that have already appeared in [4] continue to be used in this paper, and the proposed improvements and extensions mainly involve the following three steps:

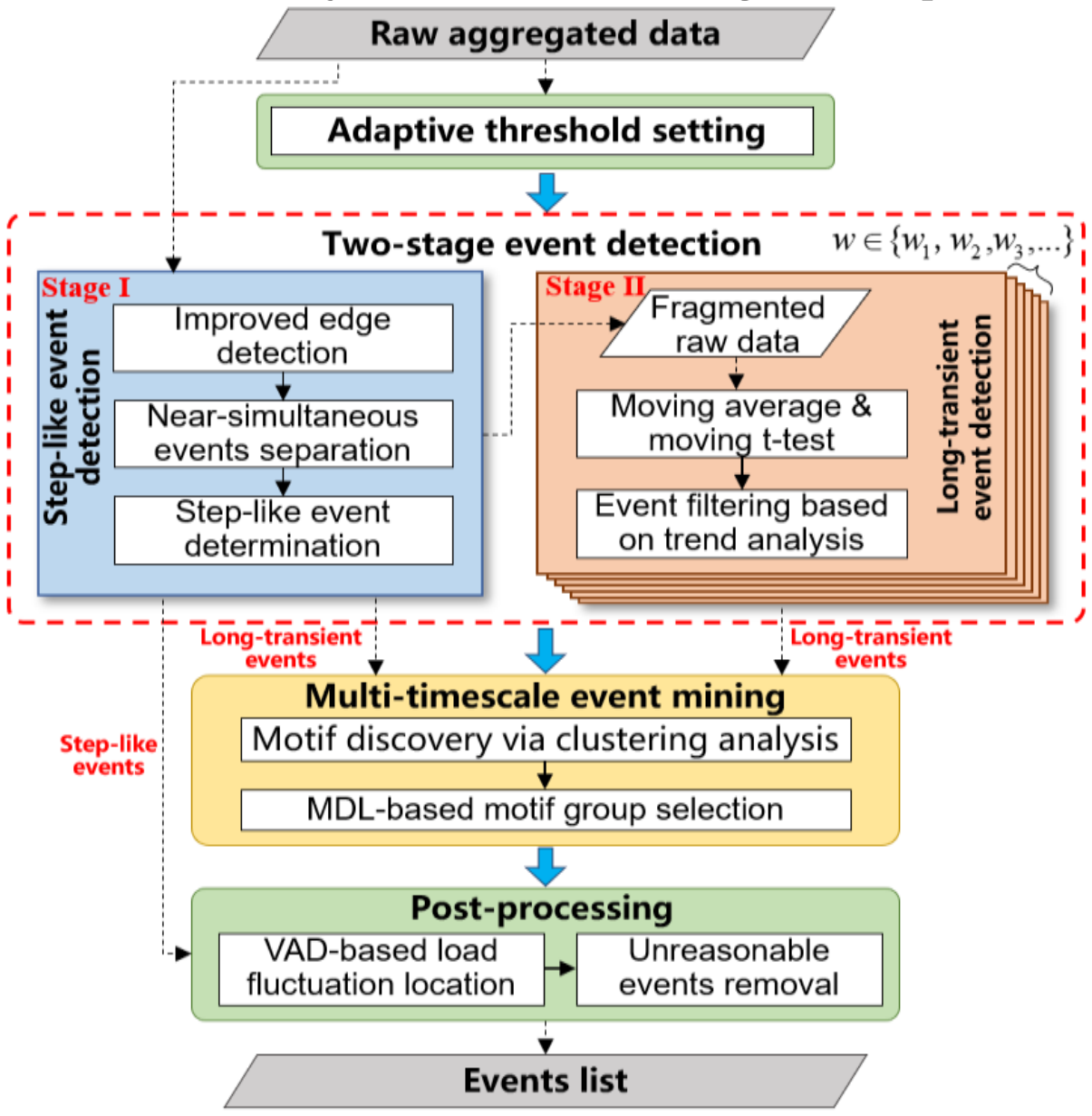

Fig. 4. MDL-based multi-timescale load event detection framework.

1) When performing the stage II of event detection, on one hand, different from the fixed window length setting of [4], long-transient event detection based on variable-length sliding window is performed on each aggregated power segment other than the segments with the step-like events detected in the stage I. On the other hand, a new step of event filtering is added to screen out the unreasonable events by power time series trend analysis of the detected event segments.

2) In order to choose among the temporally overlapping long-transient event results from different stages and different windows, a totally new multi-timescale event mining step has been added between the two-stage event detection and post-processing steps already included in [4]. The long-transient (non-step-like) events detected by the improved edge detection method in Stage I and the variable-length sliding window method in Stage II are fed together into the multi-timescale event mining step to, from which the recurring event patterns (motifs) are first discovered through unsupervised clustering analysis, and then the optimal selection is made from the competing event (motif) groups based on MDL principle.

3) In the post-processing step, different from the simple discriminant rule in [4], the unreasonable events are screened out according to the background load fluctuation characteristics of power signals. To this end, inspired by the voice activity detection (VAD) method from the field of speech signal processing, a load fluctuation segment location method is proposed to locate the fluctuation segments of aggregated power.

### *A. Long-transient event detection using variable-length sliding window*

In the second stage of event detection, this paper takes each aggregated power segment other than those transient segments of step-like events detected in Stage I as the detection object, and still implements the window-based method combining moving average and moving t-test in [4]. The criteria for determining the occurrence of events is shown in (1), and due to limited space, the method for determining the starting and ending times of the event will not be repeated here.

$$\left|\bar{P}^{(t)}_{\text{after}}-\bar{P}^{(t)}_{\text{before}}\right| > D_{\text{th}}(t) \wedge \frac{\left|\bar{P}^{(t)}_{\text{after}}-\bar{P}^{(t)}_{\text{before}}\right|}{\sqrt{\left((\xi^{(t)}_{\text{before}})^2+(\xi^{(t)}_{\text{after}})^2\right)/w}} > t^{\varphi/2}_{2w-2} \qquad (1)$$

where, at each time point $t$, $\bar{P}^{(t)}_{\text{before}}$ and $\bar{P}^{(t)}_{\text{after}}$ are respectively the average power in the window with the length of $w$ before and after it, $\xi^{(t)}_{\text{before}}$ and $\xi^{(t)}_{\text{after}}$ are respectively the standard deviations of the power time series in the corresponding windows. $D_{\text{th}}(t)$ is the detection threshold calculated for this time in the adaptive threshold setting step [4], $t^{\varphi/2}_{2w-2}$ is the critical value of t-test statistic that can be obtained by reading the t distribution table, and its freedom degree is $2w-2$.

In this paper, instead of the single fixed window for event detection taken in [4], variable window lengths $w$ are used to perform multiple event detection for the same power segment, and the detection results are further selected by the subsequent multiscale event mining step. The window length is taken from a predefined discrete set for a given sampling frequency. It is worth noting that although the improved edge detection algorithm in Stage I does not use sliding window, according to its detection principle, it can be regarded as a special window-based method with the window length of one sample.

The events detected using long sliding windows often exhibit the form of long power time series, which makes it tend to include some unreasonable events that clearly do not satisfy the fundamental shape features of the events, as shown in Fig. 5. To reduce the workload of subsequent event mining steps and prevent these unreasonable events from being selected as final results, in this paper, after finishing event detection under each window length, trend analysis on the power time series of all detected events is carried out to identify and remove those unreasonable events.

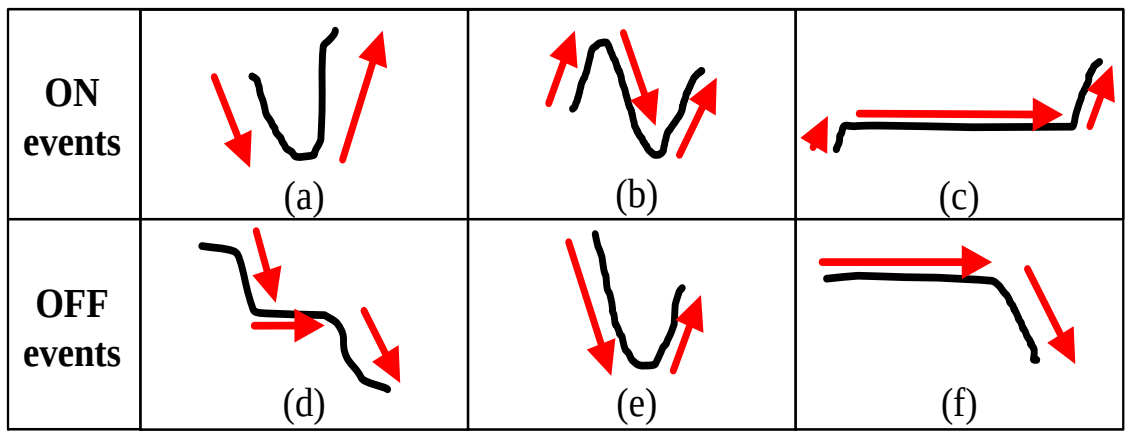

Fig. 5. Typical unreasonable events: (a) Starting with a "decreasing" trend; (b) "increasing-decreasing" trend, and the decrease is greater than the increase; (c) Overlong "steady" trend; (d) "decreasing-steady-decreasing" trend; (e) "decreasing-increasing" trend; (f) Overlong "steady" trend.

Specifically, firstly, the power time series of each event is divided into several segments using the piecewise linear representation (PLR) [4], the power changing trend of each segment is described as "increasing", "decreasing", or "steady". Thus, the overall power changing trend of the event can be represented by the combination of the trends of its all segments. Herein, the key lies in determining the "steady" trend segment. For a segment of the power time series $\{P(t_1), P(t_2), ...., P(t_n)\}$, the power difference $\Delta P_{\text{seg}} = P(t_n) - P(t_1)$ between the start and end points and the slope of the fitted straight line $k_{\text{seg}} = \Delta P_{\text{seg}} / (t_n - t_1)$ is calculated. If (2) holds, the segment can be determined to be a "steady" trend segment.

$$\begin{aligned}(|k_{\text{seg}}| < k_{\text{th_1}}) \vee (|\Delta P_{\text{seg}}| < \Delta P_{\text{th_1}}) \\ \vee ((|k_{\text{seg}}| < k_{\text{th_2}}) \wedge (|\Delta P_{\text{seg}}| < \Delta P_{\text{th_2}}))\end{aligned} \qquad (2)$$

where $k_{\text{th_1}}, k_{\text{th_2}}$ and $\Delta P_{\text{th_1}}, \Delta P_{\text{th_2}}$ are the defined power slope and power difference thresholds respectively, subject to $k_{\text{th_1}} < k_{\text{th_2}}$ and $\Delta P_{\text{th_1}} < \Delta P_{\text{th_2}}$.

When the time span of a "steady" trend segment is larger than a certain time threshold $\Delta t_{\text{steady}}$, it is considered as overlong. After detecting the "steady" trend segments, the "increasing" and "decreasing" trend segments can be identified by the sign of the power difference between the start and end points of each segment. Finally, the events are screened according to their overall power changing trends, and the unreasonable events as shown in Fig. 5, which do not fit the event fundamental shape features, are removed from the event list.

### *B. MDL-based multi-timescale event mining*

For one certain appliance transient, the event detection results under different window lengths often show different time-scale characteristics. For example, our proposed two-stage detection methods are used to detect the events of intermediate state transition processes of air conditioner (AC) shown in Fig. 6 (a), and the results shown in Fig. 6 (b) can be obtained. The set of windows selected for the window-based method is herein $w = \{5, 10, 15, 20, 25, 30\}$ (seconds), and the window length corresponding to the result of the improved edge detection method is 1 (second). In Fig. 6 (b), each event segment is covered with the same color, and adjacent events are distinguished by different colors of red and green. In this step, we will use the context information in the aggregated load data to mine the multi-timescale events detected under different windows, and ultimately select the detection results corresponding to the complete transient process based on the MDL principle.

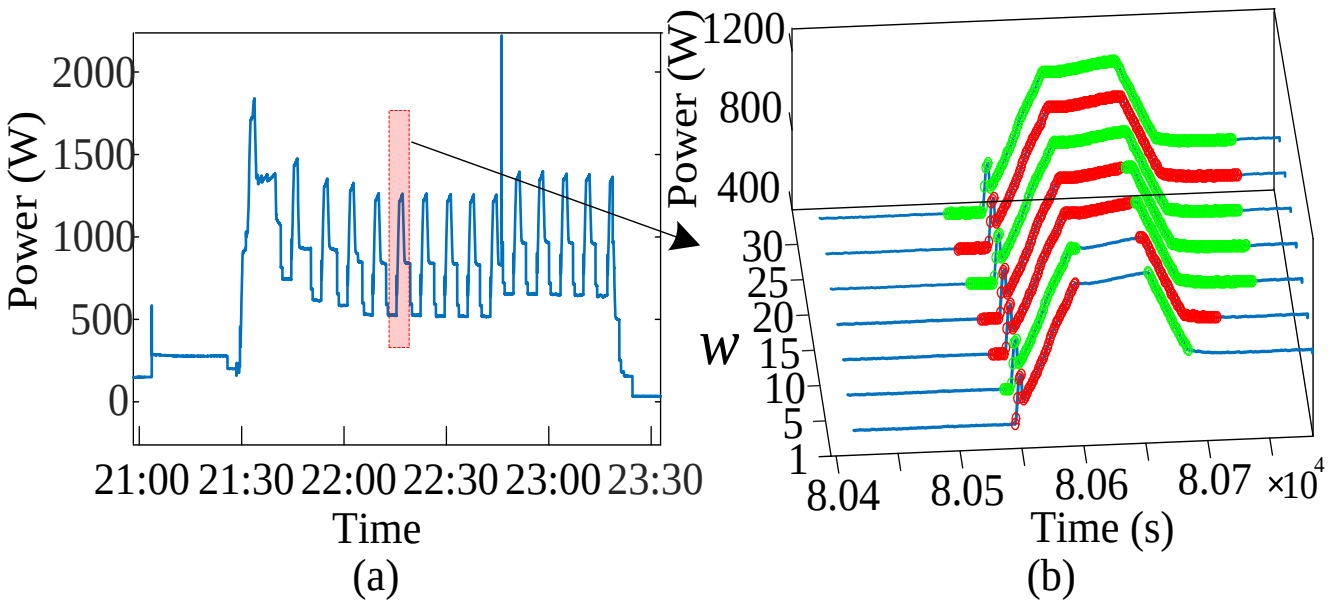

Fig. 6. Schematic diagram of event detection results under different windows. (a) Power waveform segment containing an AC, (b) Event detection results.

#### *1) Motif discovery via clustering analysis*

Obviously, the recurring events more likely correspond to the actual appliance transient processes. On one hand, the appliances are usually used many times during a period (e.g., a day, a week, or a month); on the other hand, the events of the same type could occur frequently for some appliances in a particular operating mode (e.g., the AC in the warm-keeping mode as shown in Fig. 6). Reflected in the aggregated load power time series, the repeated event power time series can be recognized as a frequent time-series pattern, which is often called "motif" in the field of time-series data mining [22]. Therefore, when the events detected under different window lengths are competing, we give priority to the event that can be identified as "motif", which is actually an exploitation of the context information in the aggregated load data.

Motif discovery is the process of unsupervised extraction of subsequence patterns from a long time series, and the routine often requires the examination of all possible subsequences in a time series[22]. Since the final results of event detection are only those power time-series subsequences that correspond to the transition processes of appliances (i.e., significant power changes), we consider the power time-series of events under different windows as candidate motifs, and the motif discovery is directly performed on them. In the subsequent process, the resulted motifs will be used as reference for the selection when different events are conflicting in time. It is worth noting that, our adopted event-driven motif discovery can reduce the search space of candidate subsequences of motifs, and its specific process is as follows:

Firstly, for the set $\boldsymbol{\Omega}$ of all long-transient events detected under each window length in $N$ days of data, all of them are clustered using the mean-shift algorithm [23]. To ensure that the same appliance have been used multiple times to meet the formation conditions of motif, the time span $N$ shall be long enough. In clustering, the feature vector representing each event sample is selected as $[\Delta P_{\text{evt}}^{(i)}, \Delta Q_{\text{evt}}^{(i)}, R_{\text{evt}}^{(i)}]$ in order to balance the clustering accuracy and computational complexity. $\Delta P_{\text{evt}}^{(i)}$ and $\Delta Q_{\text{evt}}^{(i)}$ are respectively the active and reactive power variations between the starting time $t_{\text{start}}(i)$ and ending time $t_{\text{end}}(i)$ of event $E^{(i)}$. $R_{\text{evt}}^{(i)} = \max(\boldsymbol{P}_{\text{evt}}^{(i)}) - \min(\boldsymbol{P}_{\text{evt}}^{(i)})$ is the range of active power time series $\boldsymbol{P}_{\text{evt}}^{(i)} = \{P(t)|_{t=t_{\text{start}}^{(i)}, t=t_{\text{start}}^{(i)}+1, \ldots, t_{\text{end}}^{(i)}}\}$ of the event, which can reflect the shape characteristics.

Secondly, each cluster of events is screened. Since the event detection results under different window lengths are likely to be repeated, there may be redundant events with time overlap in the same cluster. Therefore, screening is carried out for the events of each cluster to see if overlapping exists. Since the window-based method under long windows often detects some power samples outside the event transient segment as well, the one with the shortest time span among the overlapping events are retained, and the rest are directly removed from the event set $\boldsymbol{\Omega}$, which is updated to $\boldsymbol{\Omega}'$, and each event cluster is also updated accordingly.

Finally, each updated cluster of events from the set $\boldsymbol{\Omega}'$ is investigated. If the number of the cluster's elements is greater than the threshold $n_{\text{th}}$, it is considered that all its elements belong to the same motif. In this paper, the *j*-th motif is recorded as $M_j$, and the number of elements in the cluster corresponding to the motif $M_j$ is represented as $N_{\text{motif}}^{(j)}$.

*2) MDL-based motif group selection*

For the event set $\boldsymbol{\Omega}'$, the events that do not overlap with other events in time can be directly retained and sent to the subsequent post-processing steps. For the competing events that overlap each other in time, it is necessary to make the optimal selection from them according to certain principles. In the aggregated load power data, the union of the transient segments of overlapping events is taken to form the corresponding overlapping area. Even if the events in the overlapping area are identified as motifs, different motifs may overlap each other. As shown in Fig. 7, each yellow area represents an overlapping area. Each motif or the group of several non-overlapping motifs can be considered as an "explanation" of the aggregated load data in the overlapping area. What we need to do is to select the one that is most consistent with the data generation mechanism from the competing explanations, that is, it really corresponds to the complete transient process of the appliance. This is a typical problem of model selection.

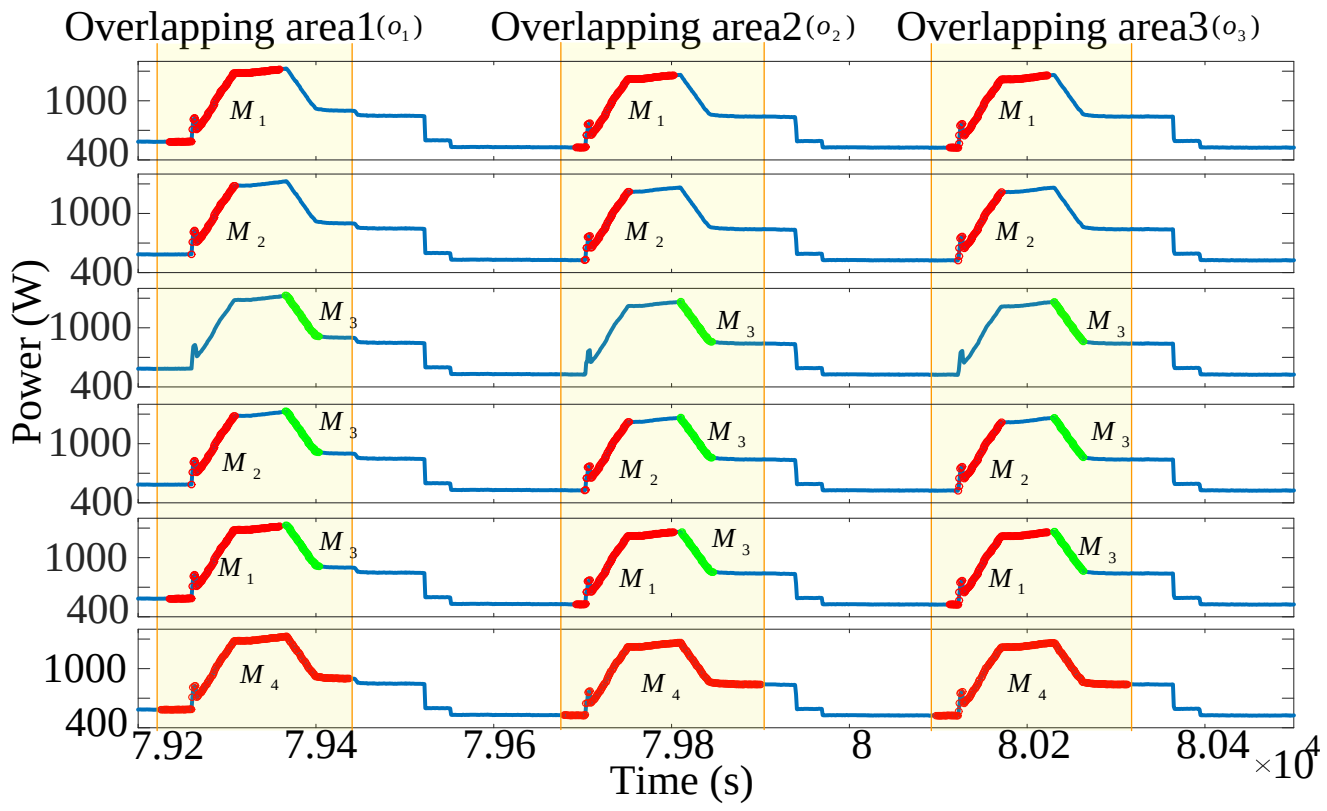

Fig. 7. Schematic diagram of overlapping areas and motif groups in the aggregated power waveform segment from Fig. 6.

We consider making decisions according to the principle of minimum description length (MDL) originated in information theory. MDL provides a general solution to the model selection problem: for a certain data $D$, given a set of hypotheses $\mathcal{H}$, or a model set, the best hypothesis $H \in \mathcal{H}$ to explain the data $D$ can be selected through (3):

$$\underset{H\in\mathcal{H}}{\operatorname{argmin}} \; L(H) + L(D|H) \tag{3}$$

where, $L(H)$ is the length, in bits, of the description of the hypothesis $H$; $L(D|H)$ is the length, in bits, of the description of the data $D$ when encoded with the help of $H$. It can be seen that the best model (hypotheses) selected by MDL is the one with the smallest overall description length of the data, that is, the model with the best compression of the original data. The regularity behind the data can be learned through MDL [22][24].

In this paper, we adopt the multi-day decision-making mechanism, that is, the final decision-making result of the day is determined by the data of the current day and the previous $N-1$ days. The specific process is as follows:

First, all possible motif groups are determined in each event overlapping area in $N$ days. Each motif group $MG$ is composed of non-overlapping motifs. For example, in each overlapping area in Fig. 7, a total of 6 motif groups are generated from 4 motifs, and each row represents a motif group. For each motif group $MG$, its description length can be defined and calculated according to (4):

$$DL(MG) = \sum_{M_j \in MG} DL(M_j) = \sum_{M_j \in MG} \frac{1}{N_{\text{motif}}^{(j)}} \sum_{E^{(i)} \in M_j} n_{\text{evt}}^{(i)} \tag{4}$$

where, the description length of $MG$ is equal to the sum of the description lengths of all motifs in the group. $DL(M_j)$ is the description length of motif $M_j$, and its value is equal to the average description length of all events belonging to the motif $M_j$. For the description length of event $E^{(i)}$, the conventional length description method [22] in bits is not adopted here. The number $n_{\text{evt}}^{(i)}$ of power sample points in the event segment is directly used as the description length, which can avoid the computation cost required for symbolizing the power time series.

Then, in all the competing motif groups, the best result is determined by pairwise comparison. The motif groups are compared in descending order of occurrences, i.e., the motif group with the highest number of occurrences is selected first and then compared with the motif group with the next highest occurrences in the overlapping area. During the comparison, the overlapping areas (denoted by the symbol $O$) in which both motif groups occur are selected for decision-making. For example, when comparing the motif group $\{M_1, M_3\}$ and $\{M_4\}$ in Fig. 7, $O = \{o_1, o_2, o_3\}$. The description length of these overlapping areas under the $MG$ representation can be defined according to (5):

$$DL(O \mid MG) = \sum_{o_j \in O} (DL(o_j) - \sum_{\substack{E^{(i)} \in MG \\ \& \ E^{(i)} \text{ in } o_j}} n_{\text{evt}}^{(i)}) + N_{\text{overlap}} \cdot N_{\text{motif}}^{MG} \tag{5}$$

where, the power time series of the event belonging to motif in the overlapping area is represented and replaced by a motif symbol (1 double data). $o_j$ denotes the $j$-th overlapping area, and $DL(o_j)$ is the raw number of power sample points in this area. $N_{\text{overlap}}$ is the number of overlapping areas in $O$, and $N_{\text{motif}}^{MG}$ is the number of motifs in $MG$.

Combining (4) and (5), the MDL estimation function shown in (6) can be calculated. Therefore, according to the MDL principle, for two competing motif groups, only the one with smaller $MDL(O \mid MG)$ is retained in the overlapping areas where they both occur.

$$MDL(O \mid MG) = DL(O \mid MG) + DL(MG) \tag{6}$$

Finally, update the occurrence frequency of each motif group and make a new round of comparison until only one motif group solution is retained in all overlapping areas. Taking Fig. 7 as an example, the final retained motif group is $\{M_4\}$. It is worth noting that if there exist other events in the overlapping area that do not overlap with the events in the selected motif groups, they are also retained.

In addition, for some overlapping areas, there may be no motif formation due to too few occurrences of some appliances, in this case, the event group contains more events will be retained. The reason is that, based on our experience, it is much easier to combine different events into the same appliance than to split a detected event into different events of different appliances in the subsequent process of appliance modeling or state identification, thus it is beneficial to keep more events.

### *C. Post-processing*

Since there is no pre-defined filtering step in the former processes, some unreasonable events tend to be detected when significant load fluctuations exist. For instance, as shown in Fig. 8, those marked in green are true events, while those in red are unreasonable events due to the fluctuations produced by a laptop. Therefore, a post-processing step is further carried out to screen all detected events with an aim to remove those unreasonable events. It consists of the following two steps.

*1) VAD-based load fluctuation segment location*

By locating the fluctuation segments of the aggregated load power, the unreasonable events can be identified and removed by using the intra-segment fluctuation characteristics. Here, we leverage the idea of voice activity detection (VAD), a technique aims to detect valid speech segments from a continuous speech stream [25]. When applied to the detection of load fluctuation, firstly, the steady-state segments, or the non-event segments are obtained based on the event detection results, as shown in Fig. 8. Herein, the power time series of the $k$-th steady-state segment is denoted as $\boldsymbol{P}_s^{(k)}$, and the first-order difference time series of all the steady-state segments are connected in series to obtain the difference power time series $\Delta \boldsymbol{P}_s = \{\Delta \boldsymbol{P}_s^{(1)}, \Delta \boldsymbol{P}_s^{(1)}, \ldots \Delta \boldsymbol{P}_s^{(1)}, \ldots\} = \{\Delta P_s^{(i)} | i = 1,2,3\ldots\}$, as shown in Fig. 9, corresponding to Fig. 8. It can be seen that the waveform has a great shape similarity with the speech signal.

Then, VAD is performed on the $\Delta \boldsymbol{P}_s$ based on the method in [25]. Specifically, a certain length of sliding window is used to scan the $\Delta \boldsymbol{P}_s$ signal non-overlappingly and the fluctuation characteristics within the window are calculated. When (7) is satisfied, the window can be determined as a fluctuation one.

$$\begin{cases} E_l = \sum_{i=1}^{N_w} (\Delta P_s^{((l-1)\cdot N_w + i)})^2 > \lambda_1 \\ R_l = \max_k (\max(\boldsymbol{P}_s^{(k)}) \text{-} \min(\boldsymbol{P}_s^{(k)})) > \lambda_2 \end{cases} \tag{7}$$

where, $E_l$ is the energy value of the signal in the $l$-th sliding window of $\Delta \boldsymbol{P}_s$, which equals to the sum of the squares of all the differential powers in the current window. $N_w$ is the window length. $R_l$ is the maximum value of power range of each original steady-state segment in the window. $\lambda_1$ and $\lambda_2$ are the thresholds.

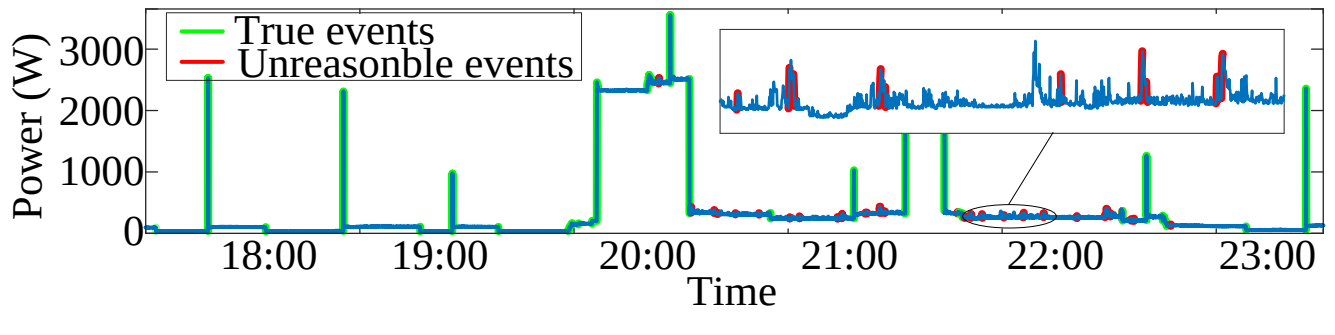

Fig. 8. A segment of aggregated power data and event detection results

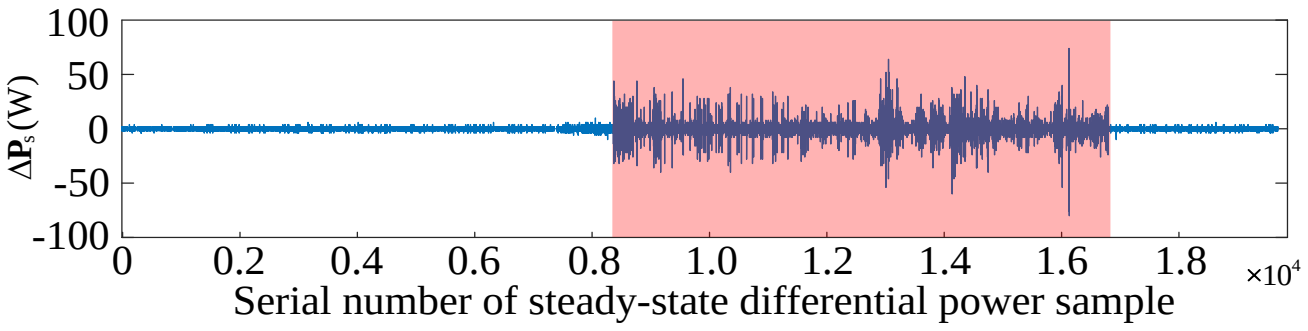

Fig. 9. Steady-state first-order differential signal corresponding to Fig. 8

Finally, the load fluctuation segment is determined and located by connecting the successive neighboring fluctuation windows, as shown in the red area in Fig. 9.

*2) Unreasonable events removal*

For each detected event $E^{(i)}$, if $\Delta P_{\mathrm{evt}}^{(i)}$ is less than the adaptive threshold, it is considered as unreasonable and removed [4]. In addition, if $E^{(i)}$ is in the load fluctuation segment and (8) is satisfied, it shall also be removed.

$$|\Delta P_{\mathrm{evt}}^{(i)}| < \mu(\{R_l\}) + \eta \cdot \sigma(\{R_l\}) \tag{8}$$

where, $\mu(\cdot)$ and $\sigma(\cdot)$ denotes average value and standard deviation respectively. $\{R_l\}$ is the sequence of $R_l$ within the time frame $W_{\mathrm{post}}$ around $E^{(i)}$ (i.e. $[t_{\mathrm{start}}(i) - W_{\mathrm{post}}, t_{\mathrm{end}}(i) + W_{\mathrm{post}}]$), And $\eta$ is the proportional coefficient (taken as 3 in this paper). Through the event screening of this step, the event false detection rate caused by load fluctuation can be effectively reduced.

## IV. Experiments

### A. Experiments settings and evaluation metrics

In our experiments, the proposed method is benchmarked with the methods in [5] and [21], and the relevant parameters are set according to their principles, which are the same as in our previous work [4].

To judge whether and to what extent the long-transient events are detected, this paper proposes an improvement on the metric $F_1^{(2)}$ based on the overlap coefficient $\varepsilon$ proposed in [4]. Where, $\varepsilon$ reflects the degree of overlapping between the transient segments corresponding to the detected event and the ground-truth event, which is defined as follows:

$$\varepsilon = \frac{\|\Delta \boldsymbol{P}_{\mathrm{overlap}}\|_1}{\max(\|\Delta \boldsymbol{P}_{\mathrm{evt}}\|_1, \|\Delta \hat{\boldsymbol{P}}_{\mathrm{evt}}\|_1)} \tag{9}$$

where, $\|\Delta \hat{\boldsymbol{P}}_{\mathrm{evt}}\|_1$, $\|\Delta \boldsymbol{P}_{\mathrm{evt}}\|_1$ and $\|\Delta \boldsymbol{P}_{\mathrm{overlap}}\|_1$ represent the sum of the absolute values of elements in active power differential time series of the ground truth event segment, the detected event segment (that overlaps with the ground truth event) and the overlapping segment, respectively.

In practical, the following two kinds of errors often occur: a long-transient event is detected as multiple events; Or the detected event covers multiple ground-truth events. In the process of calculating $F_1^{(2)}$, only the event pair with the greatest matching degree is retained to count the True Positive (TP), which is obviously pessimistic.

In this paper, we redefine the calculation method of TP based on the overlap coefficient $\varepsilon$: firstly, we construct the overlap coefficient matrix $\boldsymbol{M}_{\mathrm{OVL}}$ with dimension $n_{\mathrm{d}} \times n_{\mathrm{t}}$, where $n_{\mathrm{d}}$ is the number of detected events, $n_{\mathrm{t}}$ is the number of ground-truth events, and each element $\boldsymbol{M}_{\mathrm{OVL}}^{(i,j)}$ in the matrix is the overlap coefficient between the $i$-th detected event and the $j$-th ground-truth event; then we make a twofold judgment for each element and modify it as follows.

(a) If one element in $\boldsymbol{M}_{\mathrm{OVL}}$ is larger than a certain threshold $\rho$ (which is set to 0.8 in this paper), then the corresponding detected event is considered to be successfully matched with the ground-truth event, i.e., the operation of (10) is performed.

$$\boldsymbol{M}_{\mathrm{OVL}}^{(i,j)} \leftarrow 1, \quad \text{if } \boldsymbol{M}_{\mathrm{OVL}}^{(i,j)} > \rho \tag{10}$$

(b) If there exists a detected event that overlaps with multiple ground-truth events (and vice versa), a penalty term *penalty* (which is set to 0.1 in this paper) is applied to the element at the corresponding position, i.e., the operation of (11) is performed.

$$\boldsymbol{M}_{\mathrm{OVL}}^{(i,j)} \leftarrow \boldsymbol{M}_{\mathrm{OVL}}^{(i,j)} - penalty, \quad \text{if } \left\|\boldsymbol{M}_{\mathrm{OVL}}^{(i,\cdot)}\right\|_0 + \left\|\boldsymbol{M}_{\mathrm{OVL}}^{(\cdot,j)}\right\|_0 > 2 \tag{11}$$

Finally, TP is equal to the sum of all elements in $\boldsymbol{M}_{\mathrm{OVL}}$. In turn, the precision $Pr$, the recall $Re$, and the F1-score $F_{1-mod}$ can be calculated according to (12).

$$\begin{cases} Pr = \mathrm{TP}/n_{\mathrm{d}} \\ Re = \mathrm{TP}/n_{\mathrm{t}} \\ F_{1-mod} = 2 \cdot Pr \cdot Re/(Pr + Re) \end{cases} \tag{12}$$

Here, it is important to note that since the evaluation metrics used in this paper are different from those in our previous work [4], the test results obtained on the same test data may be different although the comparison methods and the relevant parameter settings in this paper are the same as those in our previous work [4].

### B. Results and discussion on private dataset

In this paper, the parameters of the improved and extended part of the previous adaptive two-stage event detection method[4] are set as follows: in the step of long-transient event detection using variable-length sliding window, the window length is empirically taken from a fixed set of discrete values $w \in \{5,10,15,20,25,30,60\}$ at 1 Hz frequency. $k_{\mathrm{th_1}}$, $k_{\mathrm{th_2}}$, $\Delta P_{\mathrm{th_1}}$, and $\Delta P_{\mathrm{th_2}}$ are respectively set to 0.5, 1, 10W, and 40W; and $\Delta t_{\mathrm{steady}}$ is set to 10s; In the multi-timescale event mining step, $N$ is taken as 4 and $n_{\mathrm{th}}$ is taken as 3; In the post-processing step, $N_{\mathrm{w}}$ is taken as 10, $\lambda_1$ and $\lambda_2$ are respectively taken as 50 and 5, and $W_{\mathrm{post}}$ is taken as 300. It is noteworthy that these parameters are set by analyzing the common power consumption characteristics of the typical household appliances, and are fixed for the given frequency so that they can be directly applied to different user scenarios. Table I shows the detection results of the three methods in the private dataset, and it can be seen by metric $F_{1-mod}$, which reflects the completeness of event detection, that the proposed method is significantly better than the other two methods across different user scenarios. It should be noted that Houses 1-3 in Table I are the same as our previous work [4], and House 4 is added to more fully validate the effectiveness of the proposed method in this paper.

To verify the advantages of the proposed variable-length sliding window strategy in detecting long-transient events, this paper also compares and analyzes the detection performances with the fixed single-length window detection method in [4] under different window lengths. To ensure the fairness of comparison, we also added a VAD-based post-processing step to the method in [4], and then selected different window lengths for testing on private dataset. The test results and their comparison with the proposed method are shown in Fig. 10. In House 4, the single-length window detection method achieves the best performance with a window length of 10, while the optimal window length in other scenarios is 5. It can be seen that the fixed-length window setting is difficult to adapt to different user scenarios. In contrast, the MDL-based event detection method with variable-length window can select the optimal detection result under different window lengths for each appliance transient process, thus achieving better results than the single-length window method in different scenarios.

TABLE I
EVENT DETECTION RESULTS IN PRIVATE DATASET

| | Metrics (%) | *Pr* | *Re* | $F_{1-mod}$ |
|---|---|---|---|---|
| House 1 | Proposed method | **96.26** | **95.23** | **95.74** |
| | [5] | 88.07 | 54.32 | 67.19 |
| | [21] | 95.80 | 68.30 | 79.75 |
| House 2 | Proposed method | 91.15 | **92.39** | **91.76** |
| | [5] | 85.37 | 54.12 | 66.24 |
| | [21] | **95.75** | 44.07 | 60.36 |
| House 3 | Proposed method | **88.33** | **89.98** | **89.15** |
| | [5] | 47.78 | 53.55 | 50.5 |
| | [21] | 86.87 | 36.33 | 51.23 |
| House 4 | Proposed method | **94.79** | **96.64** | **95.71** |
| | [5] | 76.28 | 39.63 | 52.16 |
| | [21] | 93.58 | 62.88 | 75.22 |

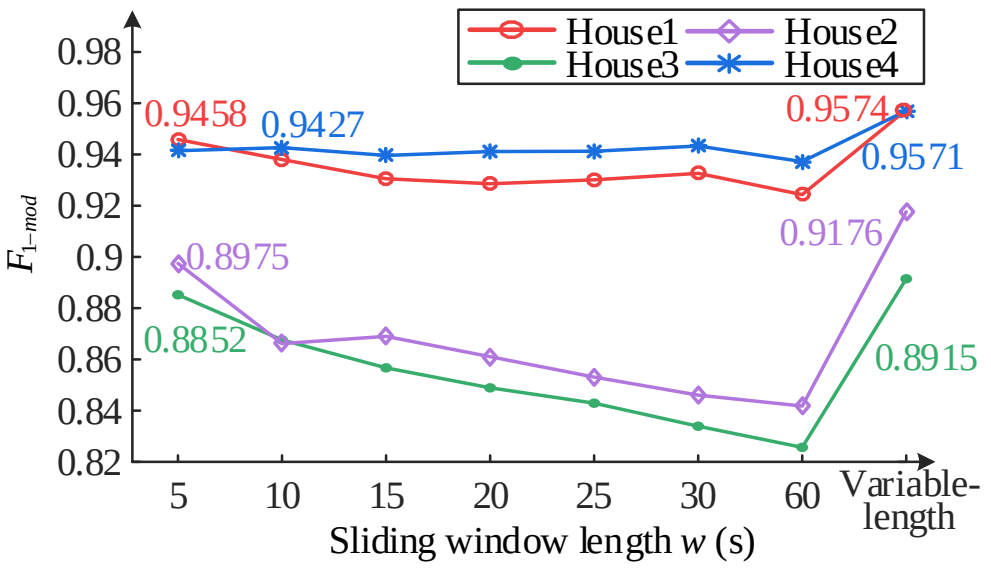

Fig. 10. Comparison between the single-length window method in [4] under different window lengths and the proposed variable-length window method

To more intuitively show the better performance of our proposed method on long-transient events, the following is a detailed analysis based on specific user scenarios. Fig. 11 (a)-(c) and (d)-(f) show, respectively, the detection results of the proposed method, [5], [21] on a segment of aggregated power data in House 3 and House 4. The whole transient segment of each event is covered by the same color, and adjacent events are distinguished by red and green colors. These two scenarios both contain long-transient events corresponding to the AC switching-on process, which can be captured completely by our method, as shown in the box in Fig. 11(a) and the boxes A, B in (d). The fixed-length window based method in [5] tends to be influenced by accidental disturbance data in the event segment, resulting in one single event splitting into different segments in Fig. 11(b) and inconsistent detection of the same appliance transient process in Fig. 11 (e). In [21], such events are missed because the window length used for clustering is smaller than the event time span, as shown in Figure 10 (c) and (f), and almost all the events detected by [21] are step-like events, which accounting for its high precision and low recall performance in Table I.

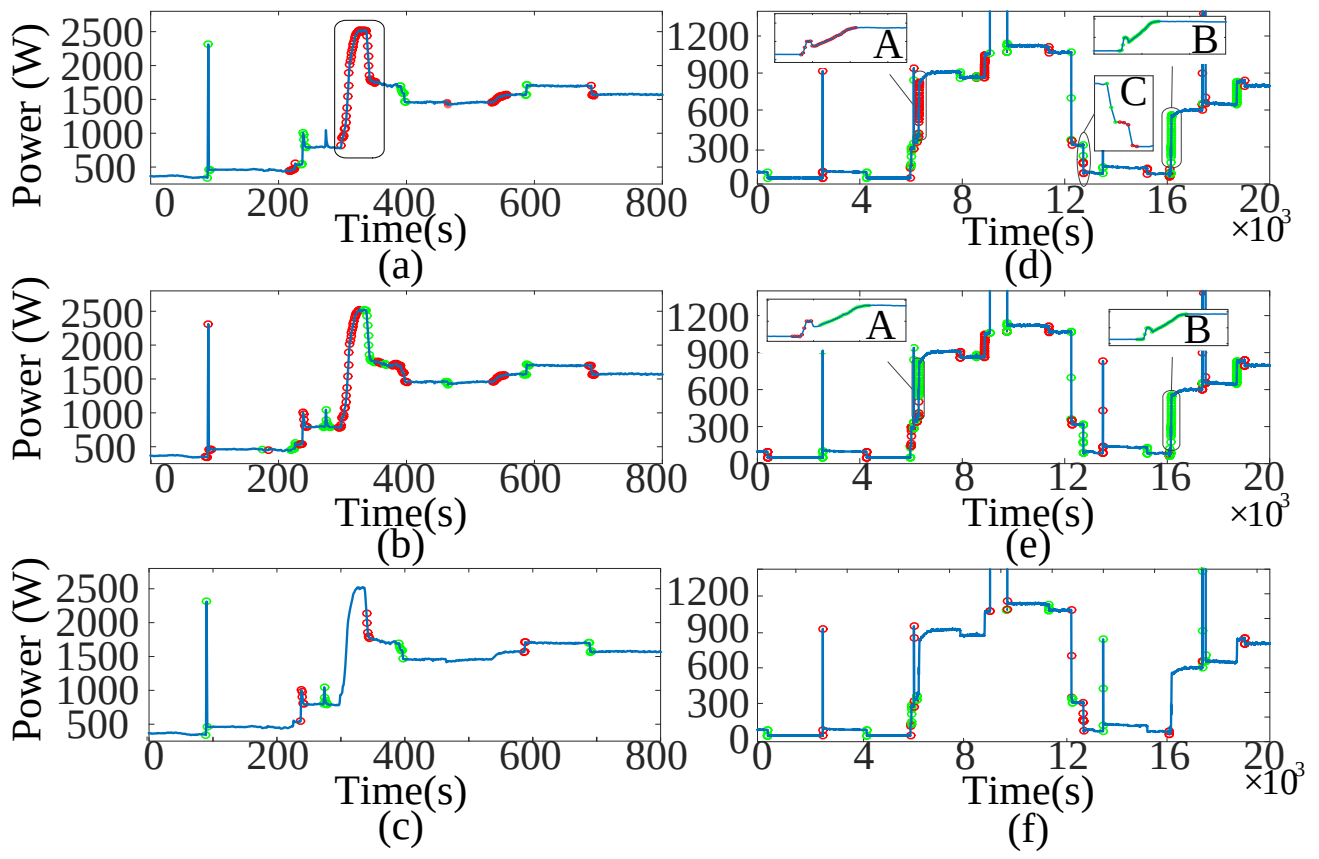

Fig. 11. Event detection results of the three methods in House 3 and 4. (a) and (d) by Proposed method, (b) and (e) by [5], (c) and (f) by [21].

Next, the role of the MDL-based multi-timescale event mining step proposed in this paper can be further verified through Fig. 12 and Fig. 13, which can also explain the reason why the AC switching-on process in Fig. 11 can be more completely detected. Fig. 12 shows the three kinds of motifs corresponding to the AC switching-on process detected in the data of House 3, which are respectively $M_1$, $M_2$ and $M_3$ corresponding to the red, green and yellow curve segments. They are obtained by clustering the events detected under different window lengths, and their occurrence distribution in the data is shown in Fig. 13. Based on the MDL principle, only the motif that can compress the data in the overlapping area to the greatest extent can be retained, which most likely corresponds to the complete transient process of the appliance (i.e., $M_1$). As can be seen, the final determination of each event segment utilizes the event information from other time periods in that scenario, and the global context information greatly improves the accuracy and consistency of detection.

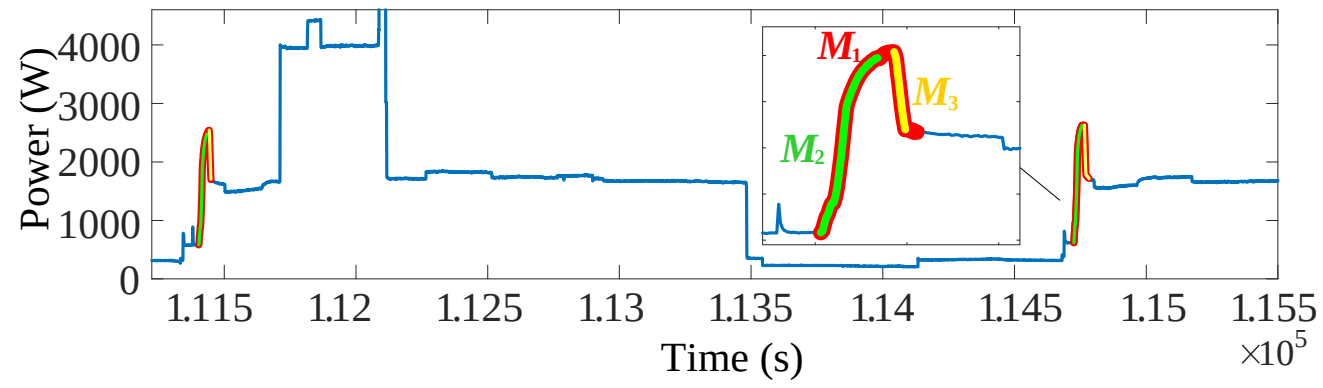

Fig. 12. Three kinds of motifs detected in House 3.

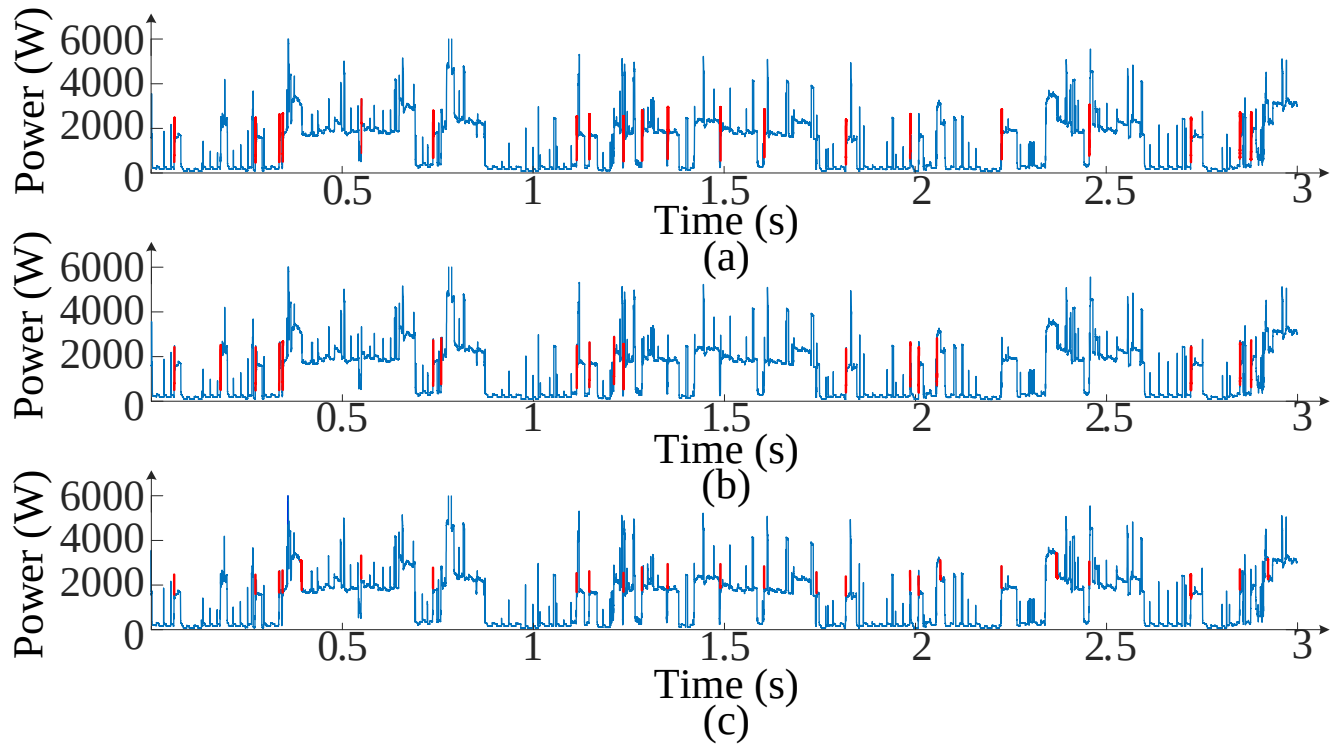

Fig. 13. Distribution of the three motifs in House 3. (a) $M_1$. (b) $M_2$. (c) $M_3$.

Finally, the ability of the proposed method to cope with load fluctuations is verified. The fluctuations that cause great interference to event detection often come from the appliances with obvious power fluctuations during operation. In Fig. 14(a)-(c), the event detection results during the AC operation in House 3 are presented, and it can be found that the power consumption presents irregular and complex fluctuations in the amplitude and time scale, which is difficult to be filtered out by a single filtering means. The method in [5] generates many false alarms as shown in Fig. 14 (b) due to the setting of the fixed detection threshold of 15W, and if the threshold is increased, a large number of events will be missed. Our method avoids the false detection of such fluctuations as events by the two steps of adaptive threshold setting and the post-processing. It is worth noting that in Fig. 14(a), some fluctuating unreasonable events with large duration still exist sporadically, indicating that the proposed method still has room for improvement. Zheng *et al* [21] avoids a large number of false alarms because it can't cluster to form any valid steady-state clusters in the fluctuation segment, as shown in Fig. 14(c). Besides, Fig.14 (d)-(f) shows the event detection results of different methods on a segment of aggregated power data containing a washing machine in House 4. The proposed method successfully locates the power fluctuation segment based on VAD in the post-processing step, as shown in the red area in Fig. 14(d), thus avoiding the false alarms by other methods as shown in box A in Fig. 14(e) and (f).

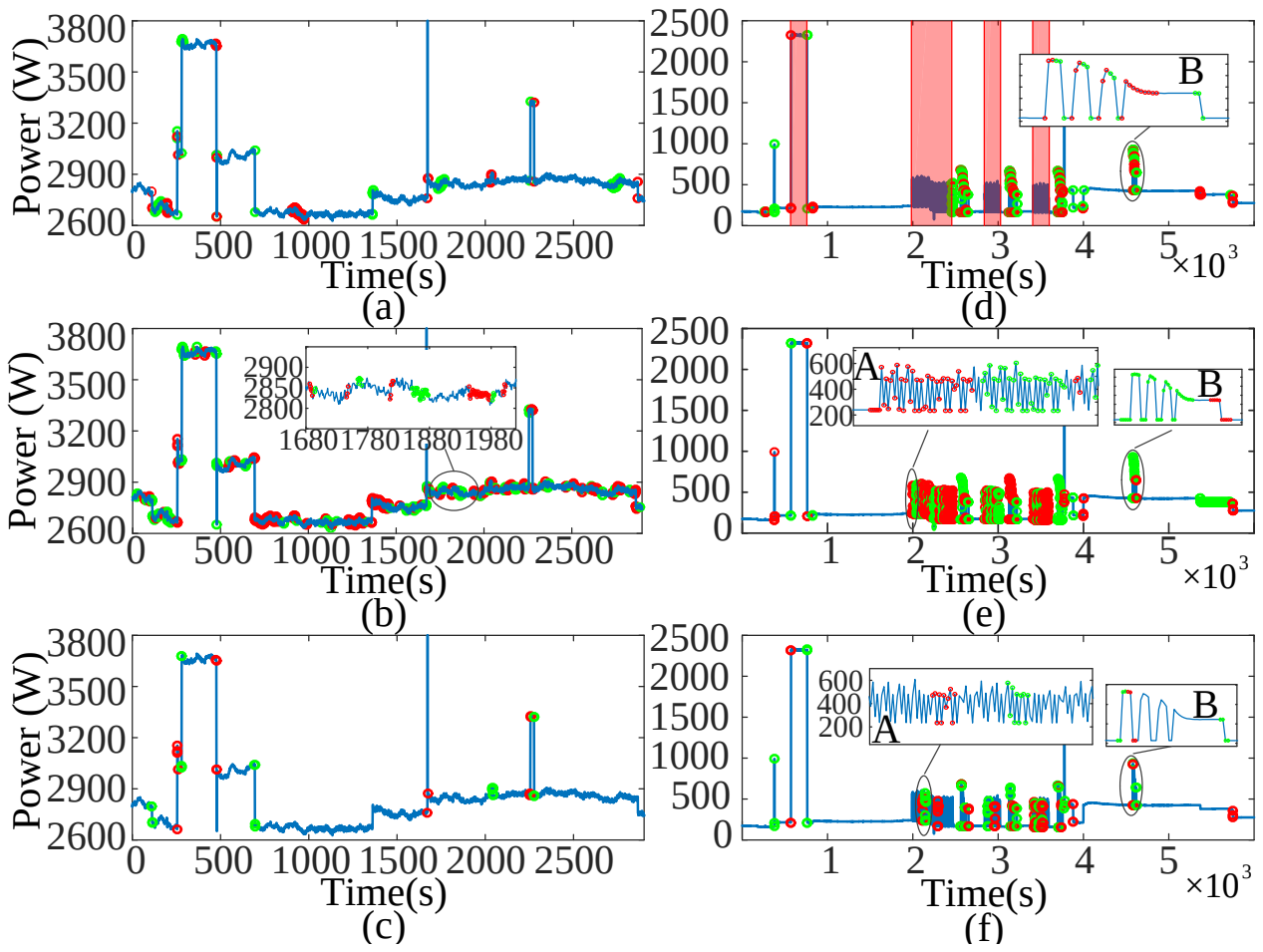

Fig. 14. Event detection results of the three methods in House 3 and 4. (a) and (d) by Proposed method. (b) and (e) by [5]. (c) and (f) by [21]

### C. Results and discussion on public dataset

In this paper, the public datasets BLUED [26] and EMBED [27] are selected for the comparison testing, and have been processed with frequency adjustment (1Hz) and manual labeling of the ground-truth events [4]. Among them, the BLUED dataset contains one user's consumption data for one week. Since the A-phase data is too simple, its B-phase data is selected for testing. EMBED dataset contains three users, and since the number of appliances connected to A and B phases in the users differs greatly, one week of testing data is selected for each user from the phase with the larger number of connected appliances (Phase A in Apt 1, Phase B in Apt 2, and Phase B in Apt 3). To ensure the fairness of testing, the parameter settings of the three event detection methods remain the same as that in Section IV-B. According to the evaluation metrics proposed in Section IV-A, the final test results are presented in Table II, and Fig. 15 shows some visualization results of the proposed method in the public dataset.

TABLE II

EVENT DETECTION RESULTS IN PUBLIC DATASET

| Metric (%) | | $Pr$ | $Re$ | $F_{1-mod}$ |
|---|---|---|---|---|
| BLUED Phase B | Proposed method | 87.52 | **84.87** | **86.17** |
| | [5] | 43.72 | 52.75 | 47.81 |
| | [21] | **91.73** | 40.05 | 55.75 |
| EMBED Apt 1 Phase A | Proposed method | 92.02 | **95.90** | **93.92** |
| | [5] | **94.69** | 85.48 | 89.85 |
| | [21] | 91.72 | 64.96 | 76.05 |
| EMBED Apt 2 Phase B | Proposed method | **99.47** | **99.33** | **99.40** |
| | [5] | 84.74 | 34.41 | 48.95 |
| | [21] | 91.92 | 70.84 | 80.01 |
| EMBED Apt3 Phase B | Proposed method | **93.22** | **97.76** | **95.43** |
| | [5] | 77.43 | 66.64 | 71.63 |
| | [21] | 80.73 | 62.16 | 70.24 |

As shown in Table II, the performance of proposed method in these two public datasets is significantly better than the other two comparison methods. It also shows good detection integrity and consistency for long-transient events, as shown in Fig. 15 (a), (b), and (c). In particular, for the event in the box in Fig. 15(b) corresponding to the AC switching-on process, its duration is as long as 5 minutes, and the power change is relatively gentle, but the cumulative power difference of the slow change part is up to 160W. It is difficult for the fixed short sliding window to capture this complete change trend. Among the three methods, only the proposed method can realize the complete detection of such events, thanks to the variable-length sliding window detection strategy.

Still taking the detection of AC switching-on events as the examples, the data in the four houses from the private dataset and from phase A of Apt 1 and Apt 3 in the EMBED dataset all contain many operation records of different kinds of ACs. According to our proposed method, the proportion of the finally determined AC switching-on events generated from different window lengths is shown in Fig. 16. It can be seen that the main window lengths (or window length distribution) corresponding to the event results vary among different houses, and the window lengths are not fixed or unique in the same house. This is because: on one hand, due to the different brands and models of ACs among different houses, the time scale and shape of the power waveforms of their events are very different; On the other hand, the event power waveforms of the same AC at different moments still vary under the influence of background noise and other factors in a given user scenario. Therefore, even for the same appliance, the optimal window length corresponding to the event that can detect it completely may not be fixed. By combining the variable-length sliding window strategy with the MDL-based multi-timescale event mining step, we optimally choose the observation window length for each event, through which the events can be detected completely and precisely, and make the contextual event detection results are as consistent as possible for the same appliance.

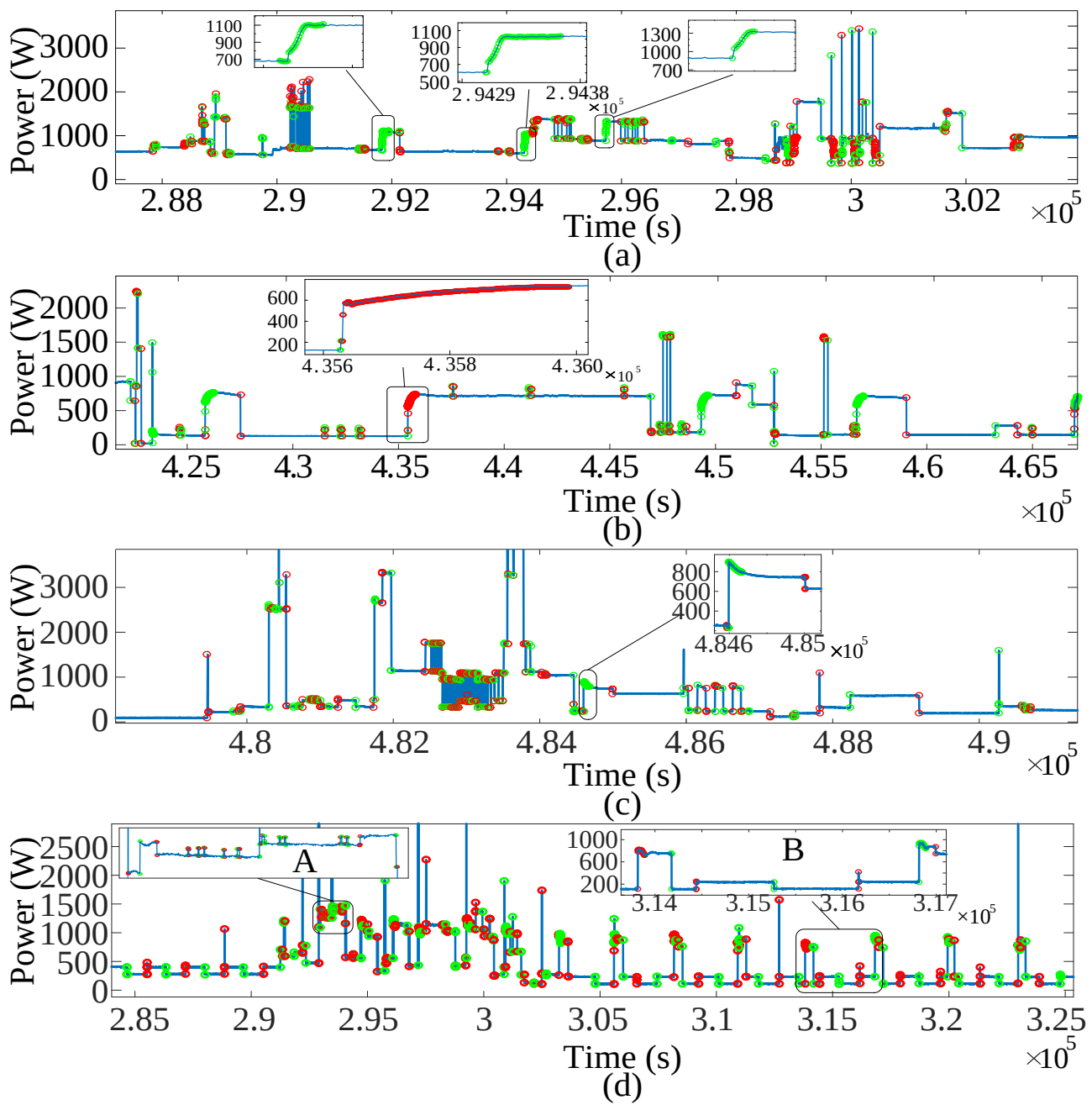

Fig. 15. Event detection results on the public datasets. (a) BLUED. (b) EMBED: Apt 1. (c) EMBED: Apt 2. (d) EMBED: Apt 3.

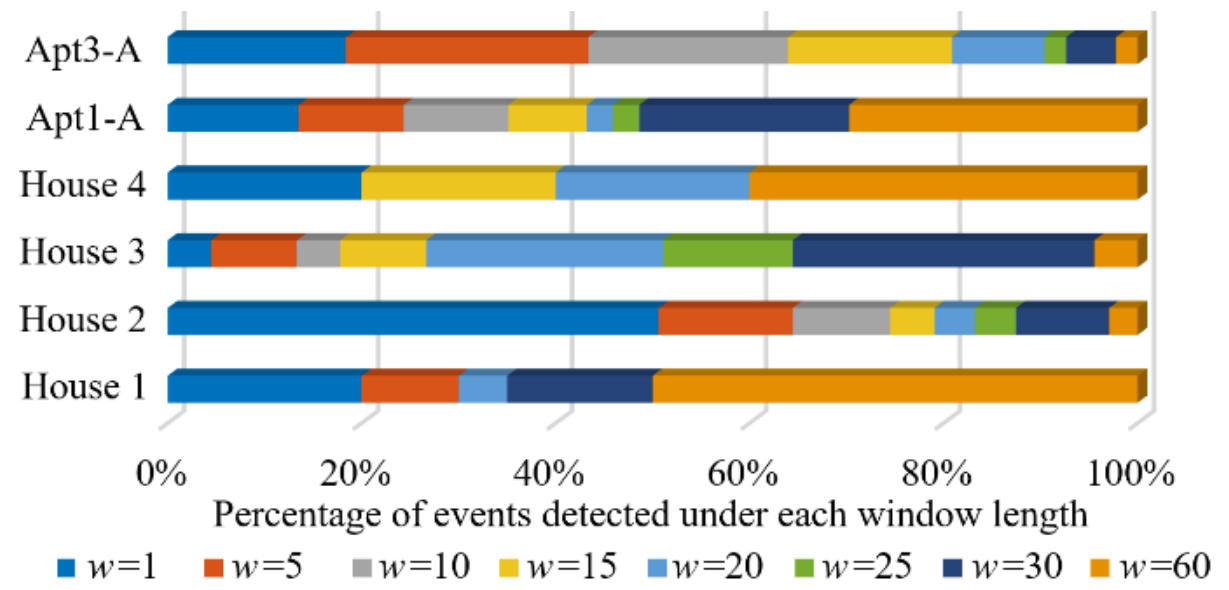

Fig. 16. Detection results of AC switching-on events in different houses.

In addition, benefiting from the adaptive two-stage event detection process, for the step-like events within each case in Fig. 15, even if the events in box A in (d) are near to occur, our method can still achieve better detection results.

## V. Conclusion

In the field of NILM, most of the existing event detection methods use fixed time parameters, which cannot adapt to the multi-timescale characteristics of events, and their accuracy tends to be affected by the load fluctuations. To cope with these problems, a multi-timescale event detection method based on MDL principle is proposed. Specifically, for the long-transient events, a novel detection scheme with variable-length sliding windows is proposed. Experiments confirm that there is always one suitable window to ensure that one certain event is detected completely. Meanwhile, by extracting and analyzing the trend components of the event power time series, the unreasonable events are screened out. Furthermore, the MDL-based multi-timescale event mining step is proposed, treating the selection of the competing event detection results under different windows as the problem of model selection, so as to select the results corresponding to the complete transient processes of appliances as much as possible, which is the first attempt in the field of NILM. In the post-processing step, a new VAD-based load fluctuation location method is proposed, and the unreasonable events wherein are removed based on analysis of the fluctuation characteristics.

Based on the newly designed evaluation metrics that can better quantify and reflect the integrity of event detection, the comparative tests on private and public datasets demonstrate that we have achieved higher event detection accuracy, owing to advantages of our adaptive two-stage event detection process for step-like events and near-simultaneous events, and the good adaptability of the proposed method to long-transient events with different time scales. In the typical residential scenarios with multiple appliances, the $F_{1-mod}$ by the proposed method can reach more than 86%. Moreover, for the long-transient events in the same scenario, the proposed method maintains good detection consistency, which is obviously better than the methods with fixed time parameter settings.

More importantly, in addition to confirming that the time-scale parameter (i.e., window length) in the proposed method can be adaptively determined based on the MDL principle, we believe that our work provides a new idea for solving the problem of key parameters adaption for event detection. If all events in the aggregated power data are accurately detected, the consistency and completeness of the detection results of the same types of events should be optimal. In this way, theoretically, the optimal description and characterization of the aggregated power profile can be achieved by using the resulting (indeed existing) power variation patterns. From the perspective of data compression, the correct event detection results can achieve the maximum compression of the aggregated power profile, corresponding to the minimum description length (in bits) of data. Therefore, for other load event detection methods, the automatic parameter selection can be achieved by solving the optimization problem with the objective of minimizing the description length of aggregated power profile based the event results detected under different parameter settings, which helps to improve their accuracy and robustness.